\documentclass{article}

\usepackage{amssymb,amsmath}
\usepackage{graphicx}
\usepackage{epstopdf}

\begin{document}

; 

\centerline{\bf A  Relativistic One  Dimensional Band Model }
\centerline{\bf with Position Dependent Mass}\vskip .2in
\vskip .2in
\centerline{ M. L.  Glasser}\vskip .2in

\centerline{Department of Physics, Clarkson University}
\centerline{Potsdam, NY 13699-5820 (USA)}
\centerline{\it laryg@clarkson.edu}\vskip .1 in
\centerline{   Departamento de F\'{\i}sica Te\'{o}rica, At\'{o}mica y \'{O}ptica
and IMUVA,}, 
\centerline{Universidad de Valladolid, 47011 Valladolid, Spain}

\vskip .3in

\centerline{\bf Abstract}\vskip .1in
\begin{quote}
In ths note a one-dimensional band model is proposed based on a periodic Dirac comb having an identical mass distribution $m(x)$ . in each unit cell. The mass function is represented as a Hermitian, non-local separable operator. Two specific cases--a constant mass model and a sinusoidal mass model--are examined. The lowest electron and positron bands for the constant mass case are similar to those for the standard relativistic Kronig-Penney model,  suggesting that non-locality has little influence. The results for the sinusoidal case are consistent with the expectation that at low wavenumber an electron ``feels" it has am average constant mass, but at high wave number, the particle "sees" the periodic mass variation and the band is distorted.

\end{quote}

\vskip .6in
\noindent
Keywords:Relativistic Dirac comb, Kronig-Penney model, enery band structure, position-dependent mass.

\vskip .2in
\noindent
PACS No.: 03.65.Fd, 03.65.Pm, 03.65.GePACS No.: 03.65.Fd, 03.65.Pm, 03.65.Ge

\newpage

\centerline{\bf Introduction}\vskip .1in

The desire for considering spatially dependent electron masses in solid state systems was expressed  as long ago as the early 1940's[1] and was made explicit through the work of Wannier, Slater, Luttinger and Kohn [2] with the development of effective mass theory in the early post-war years. Gora and Williams[3] were, it seems, the first to adapt the kinetic energy operator   to this situation, but after  noting that their original expression was not Hermitian, they proposed the  non-relativistic kinetic energy operator
$${\cal K}=-\frac{1}{4\hbar^2}[m(\vec{r})^{-1/2}\nabla+\nabla m(\vec{r})^{-1/2}]\eqno(1)$$
which was then  derived by others from various physical perspectives[4]. O.Von Roos[4]  subsequently pointed out that (1) was not unique, being  just a special case of
$${\cal K}=-\frac{\hbar^2}{4}[m^a\nabla m^b\nabla m^c+m^c\nabla m^b\nabla m^a],\quad a+b+c=-1,\eqno(2)$$
On the basis of Bargmann's theorem[5] he also  argued that (2) was unphysical and proved explicitly that the 
ambiguity  was related to a lack of Galilean invariance, i.e. observers in different inertial frames would measure different results for physical properties of such a system. Thus, Von Roos maintained that the very concept of position dependent mass should be avoided as was possible by returning to more fundamental principles. Nevertheless the position dependent mass (PDM) concept has been popular and (2) (usually with $c=a$) has been the basis of a plethora of calculations over the last 30 years. A representative set of papers where PDM is applied to various simple quantum systems is given in[6-12].

In the relativistic case, though Bargmann's theorem, that physical wave packets cannot be constructed from components  corresponding to different masses, still cannot be avoided, at least it can be argued that non-uniqueness is less of a problem since the mass  in the Dirac equation occurs simply as a linear operator, similar to the potential[13-15].   The aim here is to   take advantage of this by introducing  a   class  of exactly solvable relativistic Kronig-Penney  models  (i.e. where the lattice potential is a periodic Dirac comb with amplitude $v_0$) relative to a fixed frame where the mass is replaced by the non-local separable linear operator   
$$m(x)\psi(x)=m_0\sum_{l=-N}^N\mu(x-la)\int_{-\infty}^{\infty}dx'\mu(x'-la)\psi(x'),\eqno(3)$$
( $a$ is the lattice spacing, $N\rightarrow\infty$ the number of unit cells). As will be seen, to  construct a particular  model one need only specify the Fourier coefficients of $\mu(x)$, the mass profile in a unit cell. This note examines the case 

$$\mu(x)=\frac{1}{2}[(1+b)+(1-b)\cos(2\pi x/a)],\quad (0<b<1)\eqno(4)$$
illustrated in Fig.1 for $b=.1$ 

\begin{figure}[htbp]
\centering
\includegraphics[width=6cm]{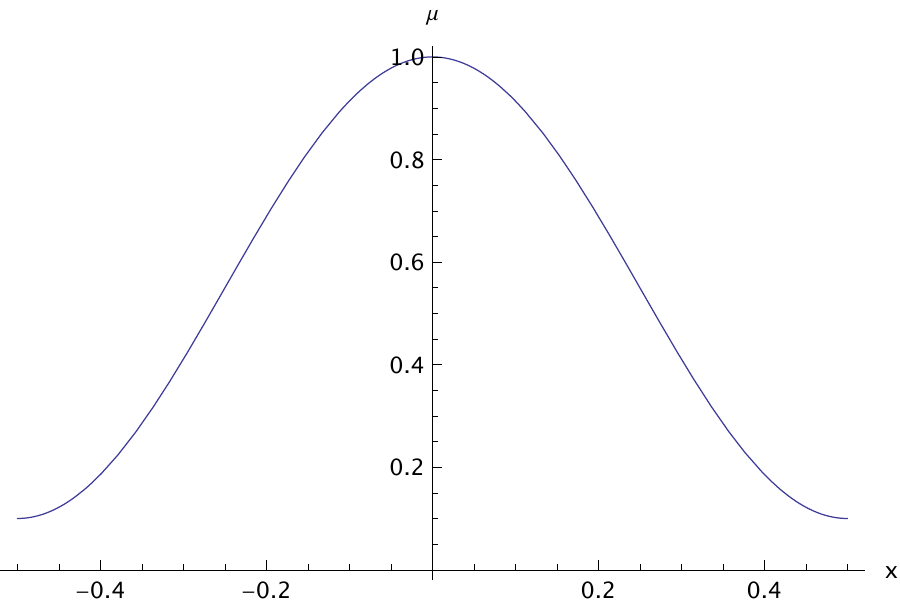}
\caption{Unit cell mass profile for $b=.1, a=5$.}
    \label{fig1}
    \end{figure}
    
 The Fourier coefficients  for (4) are

 $$\hat{\mu}_n=   
 2\frac{2\pi^2(b+1)-a^2bk_n^2}{k_n(4\pi^2-a^2k_n^2)}\sin(k_n a/2)\eqno(5)$$
$k_n=k+K_n$ and $K_n=2\pi n/a $ is the n-th reciprocal lattice vector.. The constant mass case corresponds to  $b=1$.

We present the details of the calculation in the next section, followed  by the numerical examination of the specific case  (4); the results are discussed in the concluding section.
\vskip .2in
\centerline{\bf Calculation}\vskip .1in

The one-dimensional Dirac equation can be written
$$i\hbar c\phi_1'=[E-m(x)c^2-V(x)]\phi_2(x)\eqno(4a)$$
$$i\hbar c\phi_2'=[E+m(x)c^2-V(x)]\phi_1(x)\eqno)4b)$$
and after inserting the Bloch form of the wave function components, 
$$\phi_j(x)=\sum_nC^j_n e^{ik_nx},\eqno(5)$$
where $k$ is the crystal momentum  (for simplicity we write $k_n=k+K_n$)
one has
$$\sum_n[\hbar ck_nC^1_n+EC^2_n]e^{ik_nx}=\sum_{l,n}C_n^2 e^{ik_nla}\{v_0\delta(x-la)+u_0\mu(x-la)\hat{\mu}_n\}\eqno(6a)$$
$$\sum_n[\hbar ck_nC^2_n+EC^1_n]e^{ik_nx}=\sum_{l,n}C_n^1 e^{ik_nla}\{v_0\delta(x-la)-u_0\mu(x-la)\hat{\mu}_n\}\eqno(6b)$$
where we write  $u_0=m_0c^2$. For any integrable function $g(x)$ define

$$\hat{g}_n=\int_{-\infty}^{\infty}dx'e^{ik_nx'}g(x').\eqno(7)$$
Next, we multiply each of (6a,b) by $\exp[-ik_mx]$ and integrate over $x$, noting that
$$\int_{-\infty}^{\infty}dxe^{i(k_n-k_m)x}=Na\delta_{n,m}\eqno(8a)$$
$$\sum_l e^{ik_n-k_m)la}=N\eqno(8b)$$
to find
$$\hbar ck_mC^1_m+EC^2_m=\frac{1}{a}\sum_n[C^2_n[v_0 +u_0\hat{\mu}_n\hat{\mu}_m^*]\eqno(9a)$$
$$\hbar ck_mC^2_m+EC^1_m=\frac{1}{a}\sum_n[C^1_n[v_0 -u_0\hat{\mu}_n\hat{\mu}_m^*]\eqno(9b)$$
These can be written
$${\cal{M}}{\cal{C}}=\left(\begin{array}{c}
v_0D_2+u_0d_2\hat{\mu}_m^*\\
v_0D_1-u_0d_1\hat{\mu}_m^*
\end{array}\right)\eqno(10)$$
with
$$D_j=\frac{1}{a}\sum_nC^j_n,\quad d_j=\frac{1}{a}\sum_nC^j_n\hat{\mu}_n,\eqno(11a)$$
$${\cal{M}}=\left(\begin{array}{cc}
\hbar ck_m&E\\
E&\hbar ck_m
\end{array}\right),\quad  {\cal{C}}=\left(\begin{array}{c}
C^1_m\\
C^2_m\end{array}\right).\eqno(12b)$$

Therefore, by matrix inversion
$$C_m^1=\frac{1}{\Delta_m(k)}\{\hbar c k_m[v_0D_2+u_0d_2\hat{\mu}_m^*]-E[v_0D_1-u_0d_1\hat{\mu}_m^*]\}\eqno(13a)$$
$$C_m^2=\frac{1}{\Delta_m(k)}\{\hbar c k_m[v_0D_1-u_0d_1\hat{\mu}_m^*]-E[v_0D_2+u_0d_2\hat{\mu}_m^*]\}\eqno(13b)$$
where $\Delta_m(k)=\hbar^2c^2k_m^2-E^2.$
So, in terms of the quantities
$$A_1^r=\sum_m\frac{k_m^r}{\Delta_m(k)},\quad A_2^r=\sum_m\frac{k_m^r|\hat{\mu}_m|^2}{\Delta_m(k)},\quad A_3^r=\sum_m\frac{k_m^r\hat{\mu}^*_m}{\Delta_m(k)} \eqno(14)$$
one has
$$(1+Ev_0A_1^0)D_1-\hbar cv_0A_1^1D_2-Eu_0A_3^0d_1-\hbar cu_0A_3^1d_2=0$$
$$\hbar cv_0A_1^1D_1-(1+Ev_0A_1^0)D_2-\hbar cu_0A_3^1d_1-Eu_0A_3^0d_2=0\eqno(15)$$
$$Ev_0A_3^0D_1+\hbar cv_0A_3^1D_2-(1-Eu_0A_3^0)d_1+\hbar cu_0A_2^1d_2=0$$
$$\hbar cv_0A_3^1D_1-Ev_0A_3^0D_2+\hbar cu_0A_2^1d_1-(1+Eu_0A_2^0)d_2=0$$
Consequently, the band structure is given by the roots $E(k)$ of the $4\times4$ determinant
$${\cal{D}}(k,\eta)=\left|\begin{array}{cccc}
1+Ev_0A_1^0&-\hbar cv_0A_1^1&-Eu_0A_3^0&-\hbar cu_0A_3^1\\
-\hbar cv_0A_1^1&1+Ev_0A_1^0&\hbar cu_0A_3^1&Eu_0A_3^0\\
-Ev_0A_3^0&-\hbar cv_0A_3^1&1-Eu_0A_3^0&-\hbar cu_0A_2^1\\
-\hbar cv_0A_3^1&Ev_0A_3^0&-\hbar cu_0A_2^1&1+Eu_0A_2^0
\end{array}\right|\eqno(16)$$\vskip .1in

A word about units is appropriate here. Lengths are given in Bohr radii, $a_0$ and we set $\hbar c=1$ so 
 $\eta=E/\hbar c$, $v_0$ and $u_0$ are all reciprocal lengths.
 
 The six series $A_i^r$ can be evaluated analytically; for example:
$$A_1^0=\frac{a}{2\hbar^2c^2a\eta}\left[\frac{\sin(a\eta)}{\cos(a\eta)-\cos(ka)}\right]\eqno(17a)$$
$$A_1^1=\frac{a}{2\hbar^2 c^2}\left[\frac{\sin(ka)}{\cos(a\eta)-\cos(ka)}\right]\eqno(17b)$$

\vskip.1in

 \centerline{\bf Numerical Example}\vskip .1in

As a test case, we set: $a=1,$, $v_0=2$,$u_0=1000$ and $b=0.1$. Since the ratio $ u_0/v_0=500$ is relatively small, this model may be considered mildly relativistic. The choice $a=1$, which was selected for convenience, means that the lattice is rather dense and may emphasize   anomalies. For comparison, we also examine the constant mass version: $b=1$.

The energy levels for the lowest electron band  in each case, were found  found by plotting the determinant (16), with $k$ specified,  as a function of $\eta$ and recording the  lowest positive zero. We note that there may be zeros corresponding to vanishing energy denominators $\Delta_m(k)=0$, i.e. free, zero-mass bands, which are spurious. Lengths are measured in Bohr radii and we have  set $\hbar c=1$, so that energies have units of reciprocal length.   For  model (3) with $b=1, 0.1$, the  lowest band  $E_0(k)$,  is  shown in Fig. 2, Both bands are continuous and roughly parabolic. 
    
    In Fig.(3) we show the next higher band for the two cases. Here again, that for $b=1$ lies above the one for $b=0.1$

    \begin{figure}[htbp]
\centering 
\includegraphics[width=6cm]{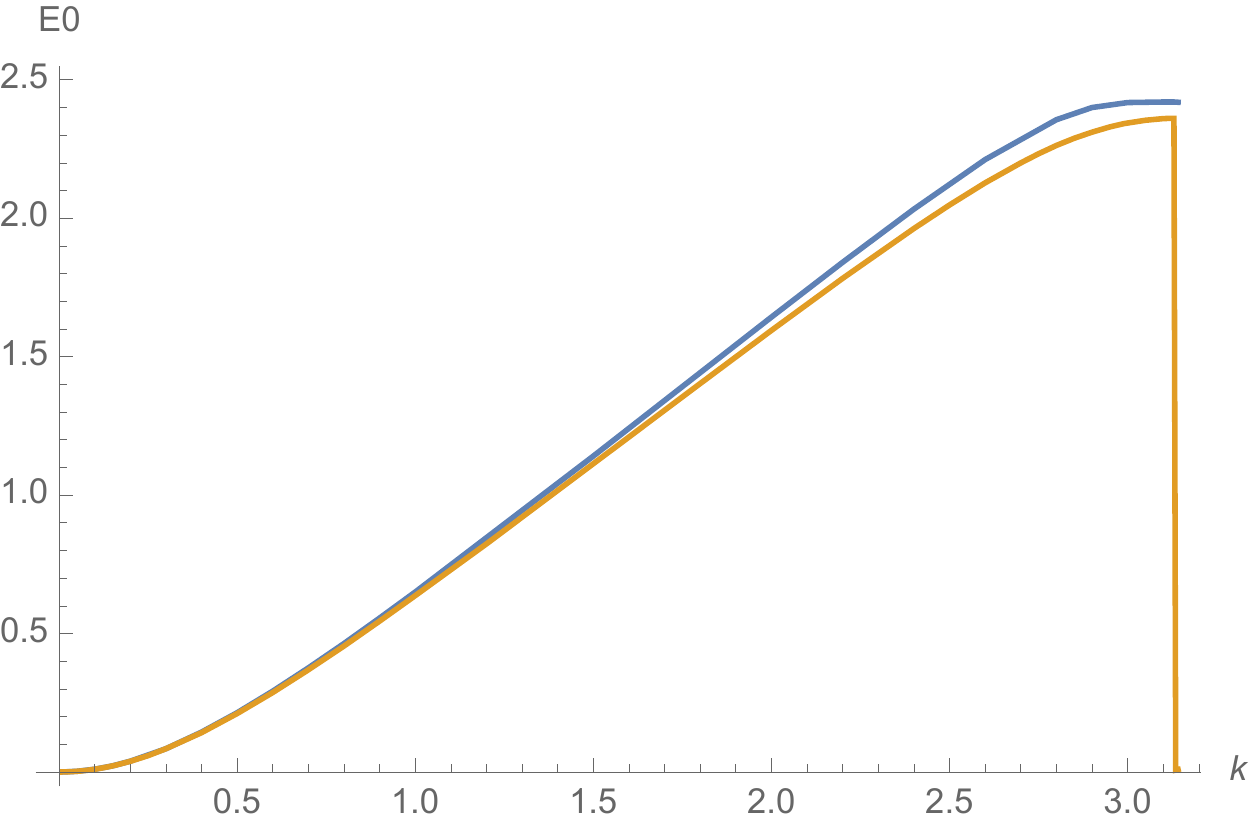}
\caption{Lowest particle  band: $a=1$, $b=.1$(lower curve), $b=1$(upper curve) $v_0=2$, $u_0=1000$.}
    \label{fig3}
    \end{figure}
    
    \vskip .1in

\centerline{\bf Discussion}\vskip .1in

Since in the constant mass case $b=1$ the lowest band resembles that for the ordinary relativistic Kronig-Penney model [16,17], it appears that the non-local nature of the mass produces no anomalies. However, at the edge of the BZ, $k=\pi/a$, the energy is driven toward zero, as indicated by a vertical line. The relative nature of the two band structures in Fig.2 can be heuristically explained as follows: At low wave vector $k$, for the sinusoidal case the particle has long wave length and "senses" only tan average mass, so the bands nearly merge, but separate at higher $k$ value where the particle wave length is small enough that the mass variation  is detectable. In. which case the energy is depressed. At the zone boundary the particle and mass wavelengths are in resonance and the particle velocity is reduced.

    \begin{figure}[htbp]
\centering
\includegraphics[width=6cm]{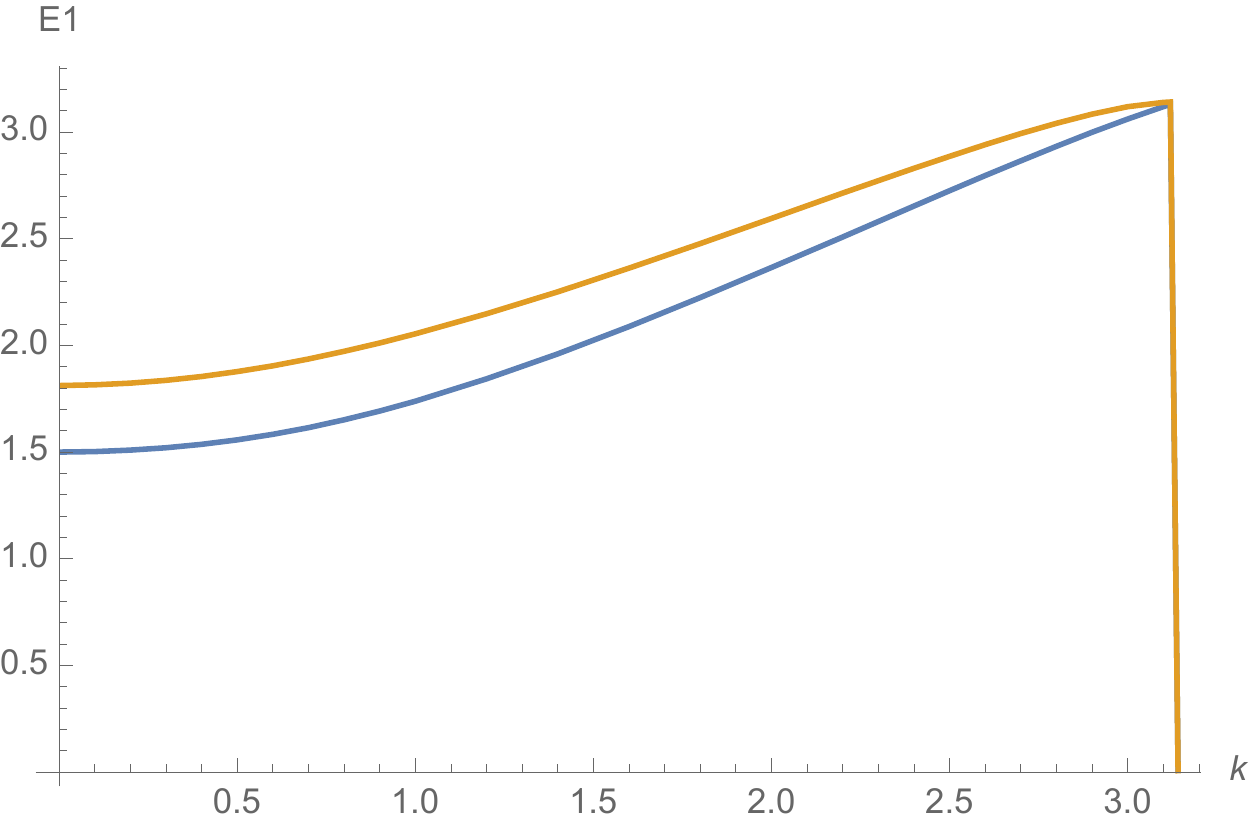}
\caption{ First excited particle  band: $a=1$, $b=.1$(initially upper curve), $b=1$( lower curve) $v_0=2$, $u_0=1000$.}
    \label{fig4}
    \end{figure}

As pointed out in [18] the upper and lower components of the two-component spinor here are particle and anti-particle wave functions, resp. Hence the negative roots of (16) correspond to positron energies. The fundamental bands of the two cases considered here are shown in Fig.4.  

In conclusion, a version of the relativistic Kronig-Penney model has been constructed which may be suitable for exploring the effects of variation of effective masses in semiconductor structures.  It incorporates the unusual replacement of the mass as a non-local operator, but this does not appear to introduce anomalies in the band structure  where the mass has purely sinusoidal variation. In a future report we hope to consider the more usual case where the mass is constant but its value varies from cell to unit cell. Finally, it is worth pointing out that the model is easily extended to two and three dimensions.

    \begin{figure}[htbp]
\centering
\includegraphics[width=6cm]{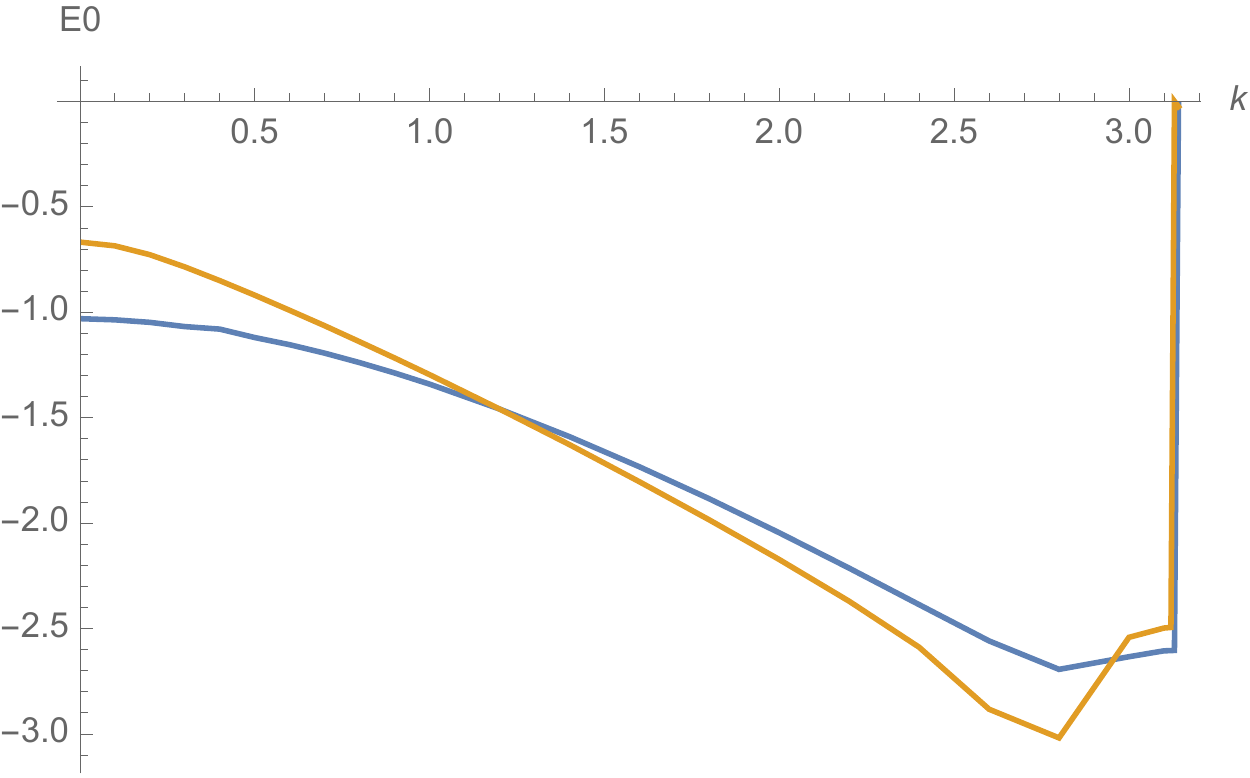}
\caption{ Antiparticle  band: $a=1$, $b=.1$(initially upper curve), $b=1$(initially lower curve) $v_0=2$, $u_0=1000$.}
    \label{fig4}
    \end{figure}
    
    \vskip .2in

 \    
 \newpage
  
  \noindent
  {\bf Acknowledgements}\vskip .1in
  The author thanks Profs. L.M. Nieto and M. Gadella for introducing him to this problem and many helpful suggestions.
He also   acknowledges  the financial support of MINECO (Project MTM2014-57129-C2-1-P) 
and Junta de Castilla y  Leon (VA057U16)).
  \vskip .1in

  \centerline{\bf References}\vskip .1in
    
    \noindent
    [1] L. Pekar,, J. Phys.(USSR){\bf 10}, 431 (1946).
    
    \noindent
    [2] G.H. Wannier, Phys.Rev.{\bf 52},192(1937); J.C. Slater,Phys.Rev.{\bf 76}, 1592(1949); J.M. Luttinger and W. Kohn, Phys.Rev.{\bf 97},869(1955).
    
    \noindent
    [3] T. Gora and F. Williams, Phys. Rev.{\bf 177},1179(1969).
    
    \noindent
    [4] O. Von Roos, {\it  Position Dependent Effective Masses in Semi-conductor Theory}, Phys. Rev.{\bf B27},7547-7552(1985); A. Trabelsi, F. Madouri and A. Matar, {\it Classification Scheme for Kinetic Energy Operators with PDM}, ArXiv:1302.3763v1(2013).
    
    \noindent
    [5] F.A. Kaempffer,{\it Concepts in Quantum Mechanics},[Academic Press. New York (1965); Appendix 7]

      \noindent
    [6]  J-M. Levy-Leblonde, {\it Elementary Quantum Models with PDM}, Eur. J. Phys. {13},316-318)1992)
    
    \noindent
    [7] R. Koc and H. T\"ut\"unc\"ulen, {\it Exact Solution of the PDM Schroedinger Equation by Supersymmetric Quantum Mechanics}, ArXiv:quant-phys-0410088v1(2004).
    
    \noindent
    [8]  S. Cruz y Cruz, J. Negro and L.M. Nieto,{\it On PDM Harmonic Oscillators}
    ,J. Phys. Conference Series {\bf 128}, 012053 (2008).
    
    \noindent
    [9] S. Cruz y Cruz and O. Rosas-Ortiz,{\it  PDM Oscillators and Coherent States},ArXiv:0902.2029v1(2009).
    
    \noindent
    [10]  Ju Guo-Xing, CAI Chang-Ying and  REN Zhang-Zhou,{\it Generalized Harmonic Oscillators and the Schroedinger Equation with PDM}, Comm. Theor. Phys.(Beijing){\bf 51}, 797-802(2009).
    
    \noindent
    [11] H. Panahi and Z. Bakhshi,{\it Solvable Potentials with PDM and Constant Mass}, Acta Physica Polobica {\bf B41},1-21(2010)
    
    \noindent
    [12]  S. Cunha and H.R. Christiansen,{\it Analytic Results in the PDM Schroedinger Equation}, ArXiv:1306.0933v1(2013).
    
    \noindent
    [13] A.D. Alhaidari, {\it Solution of the Dirac Equation with PDM in the Coulomb Field},Phys. Lett.{\bf A322},72-77(2004)
    
    \noindent
    [14] O. Panella, S. Biondini and A. Andra, {\it New  Exact Solution of the One-Dimensional Dirac Equation for the Woods-Saxon Potential with the Effective Mass Case}, J. Phys, A Math. Theor.{\bf 43} 325902(2010).
    
    \noindent
    [15]  M. Eshghi and H. Mehraban,{\it Dirac Poschl-Teller Problem with PDM}m Eur.J. Sci. Res.{\bf 54}, 22-28(2011).
    
    \noindent
    [16]  F. Dominguez Adame, Amer. J. Phys.{\bf 55}, 1004 (1987)
    
    \noindent
    [17]  P. Strange,{\it Relativistic Quantum Mechanics}[Cambridge Univ. Press., (1998)] Chap. 9.
    
      \noindent
    [18] M.L. Glasser, {\it A class of one-dimensional relativistic band models}, Amer. J. Physics {\bf  51}, 938 (1983).

\end{document}